# Determination of Boiling Range of Xylene Mixed in PX Device Using Artificial Neural Networks

Ting Zhu, Yuxuan Zhu, Hong Yang
College of Software Engineering, Sichuan University, Chengdu, Sichuan 610064, China
E-mail: clotudecho@gmail.com
E-mail: zyxsgdtc@hotmail.com
E-mail: NoraYoung2013@163.com

Hao Li[*]
College of Chemistry, Sichuan University, Chengdu, Sichuan 610064, China
E-mail: lihao_chem_92@hotmail.com

*Abstract*-Determination of boiling range of xylene mixed in PX device is currently a crucial topic in the practical applications because of the recent disputes of PX project in China. In our study, instead of determining the boiling range of xylene mixed by traditional approach in laboratory or industry, we successfully established two Artificial Neural Networks (ANNs) models to determine the initial boiling point and final boiling point respectively. Results show that the Multilayer Feedforward Neural Networks (MLFN) model with 7 nodes (MLFN-7) is the best model to determine the initial boiling point of xylene mixed, with the RMS error 0.18; while the MLFN model with 4 nodes (MLFN-4) is the best model to determine the final boiling point of xylene mixed, with the RMS error 0.75. The training and testing processes both indicate that the models we developed are robust and precise. Our research can effectively avoid the damage of the PX device to human body and environment.

*Keywords*-boiling range; determination; xylene mixed; PX device; artificial neural networks; multilayer feedforward neural networks

## I. INTRODUCTION

### A. Background

Currently, the analytical approach of xylene mixed in PX device is mainly the distillation method [1]. It's a classical approach of distillation analysis, which is convenient and precise. Therefore, distillation method is widely used in analyzing the correlate matter. However, aromatic hydrocarbon samples are harmful to human body [2-3], and impact the environment seriously [4-5]. For resolving this problem, previous researches aimed at using mathematical approaches to analyze the boiling range of correlate chemical products. L.P. Wang and her co-workers [6] have successfully proved that the boiling range of the relevant chemical products. W. Cao and his co-workers [7] has discovered the correlation approaches of gas phase chromatography distillation. As for the mathematical modeling, Y.Y. Ou [8] developed a multiple linear regression model to calculate the boiling range of xylene mixed from PX device. Previous researches indicate that the correlation between the different product components can be described by mathematical method, such as linear prediction. In our study, we tried to use non-linear function to plot this relationship.

### B. Principle of Artificial Neural Networks

Non-linear model is a powerful tool to describe the physical and biological phenomenon [9-10]. Artificial Neural Networks (ANN) [11] is one of the most useful models to describe the non-linear relationships. It is a mathematical or computational model which illustrates the possibilities of improved understanding of neural systems by applying concepts from an apparently disparate field, namely electric circuits and computer science [12].

A neural network is made up of an interconnected group of artificial neurons, using a connectionist approach to process information. Most of the time, an ANN model is an adaptive system that has the ability of adapting continuously to new data and learning from experience [13-15]. Besides, the system changes its structure based on external or internal information that flows through the network during the learning phase. They are usually used to abstract essential information from data or model complex relationships between inputs and outputs.

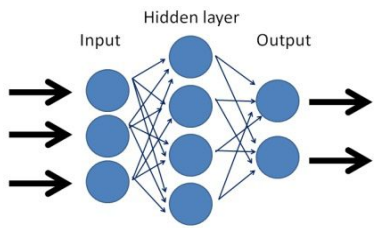

Figure 1. A schematic view of artificial neural network structure.

The main structure of the artificial neural network (ANN) consists of the input layer and the output layer (Figure 1). The input variables was introduced to the network by the input layer [16], while the network provides the response variables with predictions which stand for the output of the nodes in this certain layer. Apart from that, it includes the hidden layers. The type and the complexity of the process or experimentation usually iteratively determine the optimal number of the neurons in the hidden layers [16].

## II. Model Development

### A. Training process

22 samples were used to develop the prediction models. Thereinto, 8 groups were used as the testing set, 14 groups were used as the training set. Independent variables are made up of different main constituents of chemical product, including nonaromatic substance, toluene, ethyl benzene, p-xylene, m-xylene, isopropyl benzene, o-xylene, n-proplbenzene and C9 aromatics. Dependent variables are the boiling range of xylene mixed in PX device, containing initial boiling point and final boiling point.

To ensure the robustness of the models, two groups of experiments were done respectively, one is the determination of initial boiling point, the other is the determination of final boiling point. To find out the suitable and reasonable model of the determination, the RMS error of the testing is the main factors for consideration. In each group of experiments, we utilized linear regression, General Regression Neural Networks [17-19] (GRNN) and Multilayer Feedforward Neural Networks [19-21] (MLFN) to the establishment. Thereinto, MLFN models were experimented with different nodes, so that, the best situation of nodes could be found out. Training results are shown as follows:

TABLE I. Results of different mathematical model in determining initial boiling point.

| ANN model | Trained samples | Tested samples | RMS error |
|---|---|---|---|
| Linear prediction | 14 | 8 | 0.66 |
| GRNN | 14 | 8 | 0.23 |
| MLFN 2 Nodes | 14 | 8 | 0.37 |
| MLFN 3 Nodes | 14 | 8 | 0.37 |
| MLFN 4 Nodes | 14 | 8 | 0.55 |
| MLFN 5 Nodes | 14 | 8 | 0.54 |
| MLFN 6 Nodes | 14 | 8 | 0.25 |
| MLFN 7 Nodes | 14 | 8 | 0.18 |
| MLFN 8 Nodes | 14 | 8 | 0.32 |
| MLFN 9 Nodes | 14 | 8 | 0.21 |
| MLFN 10 Nodes | 14 | 8 | 0.21 |
| MLFN 11 Nodes | 14 | 8 | 0.39 |
| MLFN 12 Nodes | 14 | 8 | 0.20 |
| MLFN 13 Nodes | 14 | 8 | 0.52 |
| MLFN 14 Nodes | 14 | 8 | 0.50 |
| MLFN 15 Nodes | 14 | 8 | 0.56 |
| MLFN 16 Nodes | 14 | 8 | 4.01 |
| MLFN 17 Nodes | 14 | 8 | 2.42 |
| MLFN 18 Nodes | 14 | 8 | 2.24 |
| MLFN 19 Nodes | 14 | 8 | 2.63 |
| MLFN 20 Nodes | 14 | 8 | 2.49 |
| MLFN 21 Nodes | 14 | 8 | 0.98 |
| MLFN 22 Nodes | 14 | 8 | 2.19 |
| MLFN 23 Nodes | 14 | 8 | 1.55 |
| MLFN 24 Nodes | 14 | 8 | 2.18 |
| MLFN 25 Nodes | 14 | 8 | 1.70 |
| MLFN 26 Nodes | 14 | 8 | 2.49 |
| MLFN 27 Nodes | 14 | 8 | 2.31 |
| MLFN 28 Nodes | 14 | 8 | 2.59 |
| MLFN 29 Nodes | 14 | 8 | 6.68 |
| MLFN 30 Nodes | 14 | 8 | 3.17 |

Table 1 implies that the MLFN model with 7 nodes (MLFN-7) is the best model during the training process, with the RMS error 0.18.

Results of different mathematical model in determining final boiling point are shown in Table 2:

TABLE II. Results of different mathematical model in determining final boiling point.

| ANN model | Trained samples | Tested samples | RMS error |
|---|---|---|---|
| Linear prediction | 14 | 8 | 6.19 |
| GRNN | 14 | 8 | 1.38 |
| MLFN 2 Nodes | 14 | 8 | 2.13 |
| MLFN 3 Nodes | 14 | 8 | 1.76 |
| MLFN 4 Nodes | 14 | 8 | 0.75 |
| MLFN 5 Nodes | 14 | 8 | 1.89 |
| MLFN 6 Nodes | 14 | 8 | 2.36 |
| MLFN 7 Nodes | 14 | 8 | 2.29 |
| MLFN 8 Nodes | 14 | 8 | 1.53 |
| MLFN 9 Nodes | 14 | 8 | 1.04 |
| MLFN 10 Nodes | 14 | 8 | 2.15 |
| MLFN 11 Nodes | 14 | 8 | 1830.25 |
| MLFN 12 Nodes | 14 | 8 | 1.19 |
| MLFN 13 Nodes | 14 | 8 | 1.05 |
| MLFN 14 Nodes | 14 | 8 | 1.52 |
| MLFN 15 Nodes | 14 | 8 | 11.26 |
| MLFN 16 Nodes | 14 | 8 | 9.58 |
| MLFN 17 Nodes | 14 | 8 | 16.30 |
| MLFN 18 Nodes | 14 | 8 | 5.61 |
| MLFN 19 Nodes | 14 | 8 | 7.90 |
| MLFN 20 Nodes | 14 | 8 | 6.37 |
| MLFN 21 Nodes | 14 | 8 | 10.78 |
| MLFN 22 Nodes | 14 | 8 | 5.15 |
| MLFN 23 Nodes | 14 | 8 | 19.95 |
| MLFN 24 Nodes | 14 | 8 | 9.68 |
| MLFN 25 Nodes | 14 | 8 | 5.80 |
| MLFN 26 Nodes | 14 | 8 | 15.31 |
| MLFN 27 Nodes | 14 | 8 | 6.48 |
| MLFN 28 Nodes | 14 | 8 | 9.59 |
| MLFN 29 Nodes | 14 | 8 | 5.60 |
| MLFN 30 Nodes | 14 | 8 | 7.03 |

According to the results presented in table 2, MLFN model with 4 nodes (MLFN-4) possesses the lowest RMS error (0.75), which is obviously much lower than other models. Therefore, we considered that the MLFN-4 model is a reasonable model in determining final boiling point.

III. RESULTS AND DISCUSSION

A. *Determination of initial boiling point*

To describe the training process of initial boiling point, figure 2 is used to present the accuracy of the MLFN-7 model, which is shown as follows:

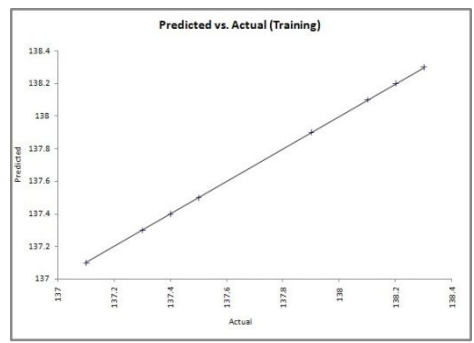

Figure 2. Comparison between predicted values and actual values of initial boiling point during training process.

Figure 2 indicates that using MLFN-7 model can describe the relationship between initial boiling point and the product components primely. The good fitting result show that the training process is stable and robust.

B. *Determination of final boiling point*

Uniformly, figure 3 is used to depict the training process of the determination of final boiling point, which is shown as follows:

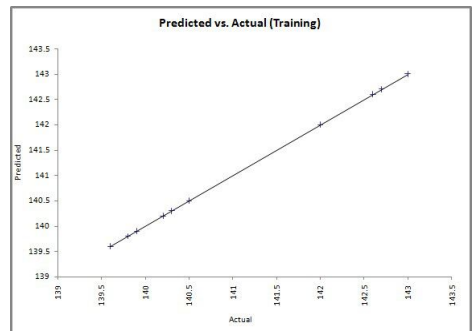

Figure 3. Comparison between predicted values and actual values of final boiling point during training process.

Figure 3 depicts the determination process of final boiling point using MLFN-4 model. Results also show that the MLFN-4 model can fit the relationship between final boiling point and product components admirably, indicating that the training process is robust and precise.

C. *Discussion*

Figure 2 to 3 depict the robustness of the ANN model in determining the boiling range of xylene mixed in PX device, indicating that the training process is accurate and reasonable. It's axiomatic that the boiling range of xylene mixed in PX device can be determined by artificial neural networks model without the complex operation of practical experiments. In addition, it's worth mentioning that Multilayer Feedforward Neural Networks (MLFN) is particularly accurate in calculating the boiling range in different nodes of networks.

IV. CONCLUSION

Determination of boiling range for xylene mixed in PX device is currently a crucial topic in the practical applications because of the recent disputes of PX project in China. In our study, we successfully established two artificial neural networks models to determine the initial boiling point and final boiling point respectively. Results show that the Multilayer Feedforward Neural Networks (MLFN) model with 7 nodes (MLFN-7) is the best model to determine the initial boiling point of xylene mixed, with the RMS error 0.18, while the MLFN model with 4 nodes (MLFN-4) is the best model to determine the final boiling point of xylene mixed, with the RMS error 0.75. The training and testing processes both indicate that the models we developed are robust and precise. Our research can avoid the damage of the PX device to human body and environment.

V. ACKNOWLEDGEMENTS